\documentclass[12pt,letter]{article}
\usepackage{graphicx}
\usepackage{slashed}
\usepackage{subfigure}
\usepackage{hyperref}

\title{The milliQan Experiment: Search for milli-charged Particles at the LHC \\ 
       (proceeding for ICHEP 2018\footnote{ICHEP 2018 SEOUL, International Conference on High Energy Physics, 4-11 July 2018, SEOUL, KOREA} )}

\author{Jae Hyeok Yoo (University of California, Santa Barbara)\\
        On behalf of the milliQan Collaboration\\
        E-mail: \href{mailto:jae.hyeok.yoo@cern.ch}{jae.hyeok.yoo@cern.ch}}
\date{}

\begin{document}

\maketitle 
Recently, a search for milli-charged particles produced at the LHC has been proposed. The experiment, named milliQan, is expected to obtain sensitivity to charges of $10^{-1} - 10^{-3}e$ for masses in the 0.1 - 100 GeV range. The detector is composed of 3 stacks of 80 cm long plastic scintillator arrays read out by PMTs. It will be installed in an existing tunnel 33 m from the CMS interaction point at the LHC, with 17 m of rock shielding to suppress beam backgrounds. In the fall of 2017 a 1\% scale "demonstrator" of the proposed detector was installed at the planned site in order to study the feasibility of the experiment, focusing on understanding various background sources such as radioactivity of materials, PMT dark current, cosmic rays, and beam induced backgrounds. The data from the demonstrator provides a unique opportunity to understand the backgrounds and to optimize the design of the full detector.  

\newpage

\section{Motivation of milliQan experiment}
No evidence for new physics at the LHC has been found. This might mean that there can be phase space that we are not exploring. The milliQan experiment~\cite{Haas:2014dda} searches for milli-charged particles (mCP) produced at the LHC. The existing detectors miss such particles due to their small energy deposits and the large background in a high luminosity environment. 

One, well motivated, mechanism to predict such milli-charged particles is to introduce a new U(1) in dark sector with a massless dark-photon ($A'$) and a massive dark-fermion ($\psi$), 
\begin{equation}
\mathcal{L}_{\textrm{dark sector}} 
= -\frac{1}{4} A'_{\mu\nu} A'^{\mu\nu} + i \bar{\psi} \left( \slashed{\partial} + i e' \slashed{A}' + i \textrm{M}_{\textrm{mCP}}\right) \psi 
  - \frac{\kappa}{2} A'_{\mu\nu} B^{\mu\nu}.
\end{equation}
where the last term shows that $A'$ and $B$ kinetically mix with the mixing parameter $\kappa$. After replacing $\slashed{A}'$ with $\slashed{A}' + \kappa B_\mu$, the coupling between the dark fermion and $B$ becomes $\kappa e'$ ($~\kappa e' \bar{\psi} \slashed{B} \psi$).  Therefore, the charge of $\psi$ can vary by the size of mixing.  
\begin{figure}[htp]
\centering
\vspace{1cm}
\subfigure[]{
  \centering
  \label{subfig:sensitivity}
  \includegraphics[width=0.85\textwidth]{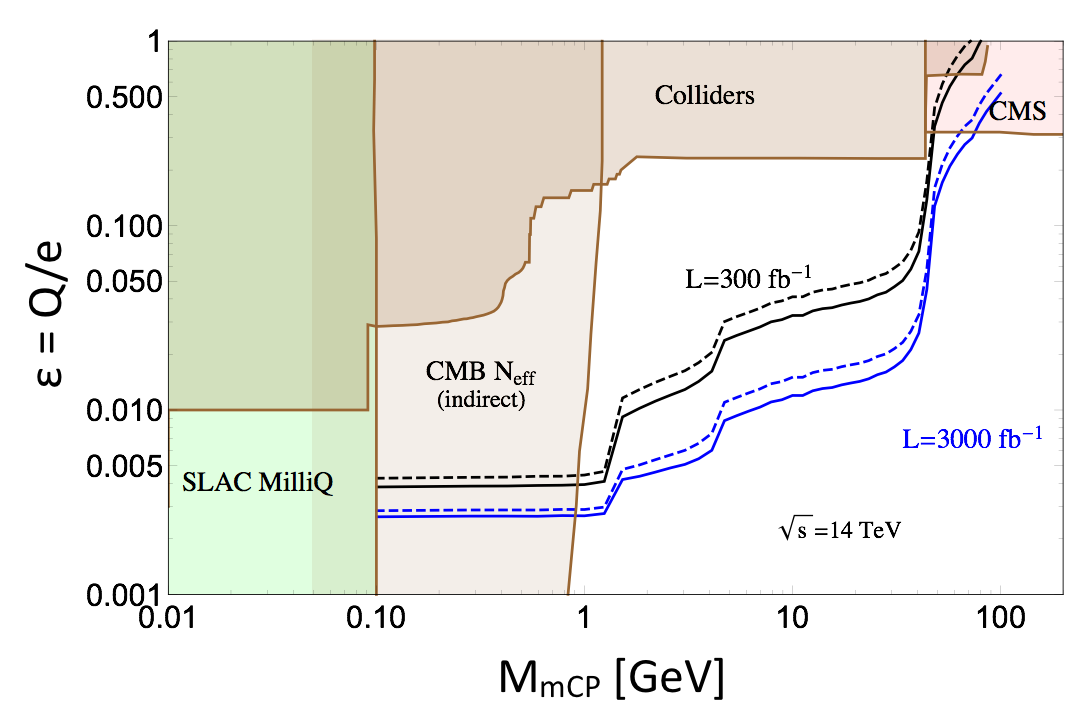}
}
\hspace{1cm}
\subfigure[]{
  \centering
  \label{subfig:concept}
  \includegraphics[width=0.65\textwidth]{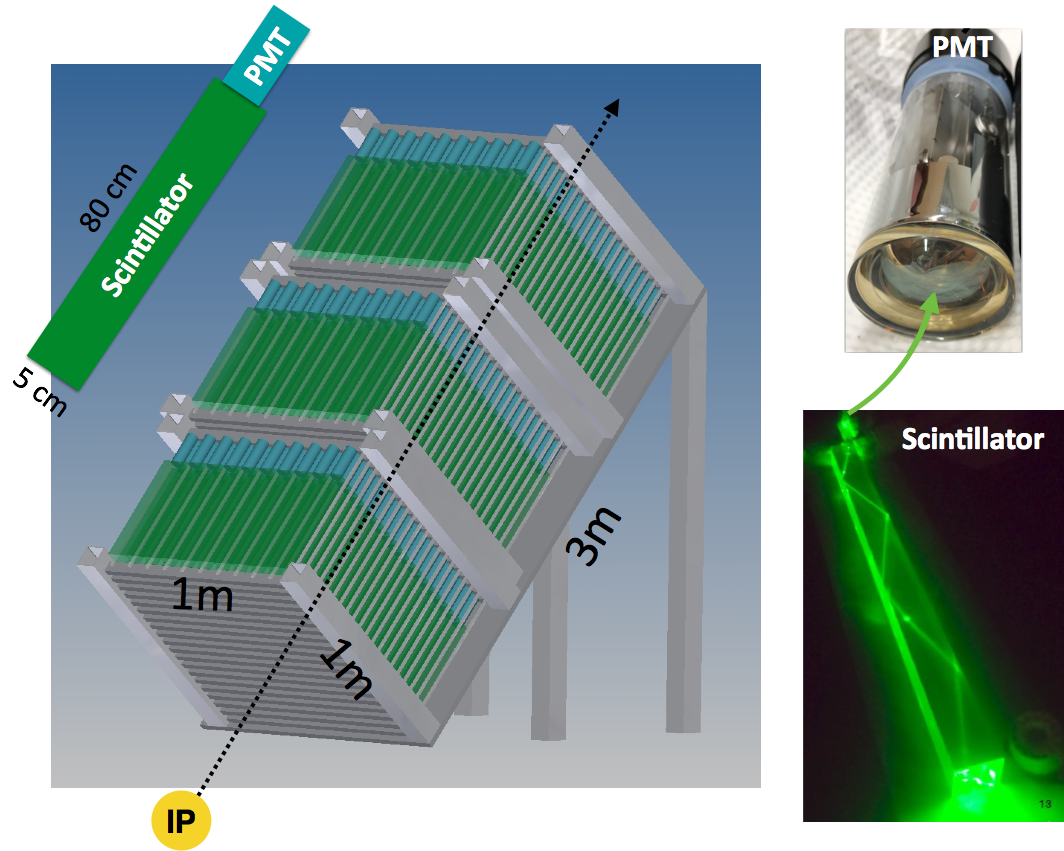}
}
\caption{ (a) Sensitivity of the existing experiments to mCP and (b) detector concept.}
\label{fig:sensitivity_concept}
\end{figure}
Figure~\ref{subfig:sensitivity} shows the sensitivity of the existing experiments. Low mass and high charge region is covered by the low energy and direct CMS/ATLAS searches, but the $M_{mCP}>1$ GeV and $Q<0.3e$ region is unexplored. The milliQan experiment provides a unique opportunity to explore this region.

\section{Detector concept and demonstrator}
The energy deposited by fractionally charged particles is proportional to $Q^2$, meaning a particle with charge down to $10^{−3}e$ has $dE/dx$ of $10^{−6}$ of a particle with $Q=1e$. Therefore, the detector should have a large, active, and sensitive area to produce signals even as small as single photon signals. As shown in Figure~\ref{subfig:concept}, The milliQan detector is composed of 3 layers of $80\times5\times5$ cm scintillator arrays pointing back to the CMS interaction point (IP). The particles from the IP should go
through all 3 layers; this feature is used to reduce random combinatoric backgrounds significantly. The light from the scintillator bars is converted and amplified by photomultiplier tube (PMT).

The detector is located in a underground tunnel at LHC P5. The distance to the CMS IP is 33 m, of which 17 m is a rock that provides shielding for particles from collisions. Since it is about 70m underground, the cosmic muons are significantly reduced.    
\begin{figure}[bp!]
\centering
\includegraphics[width=0.9\textwidth]{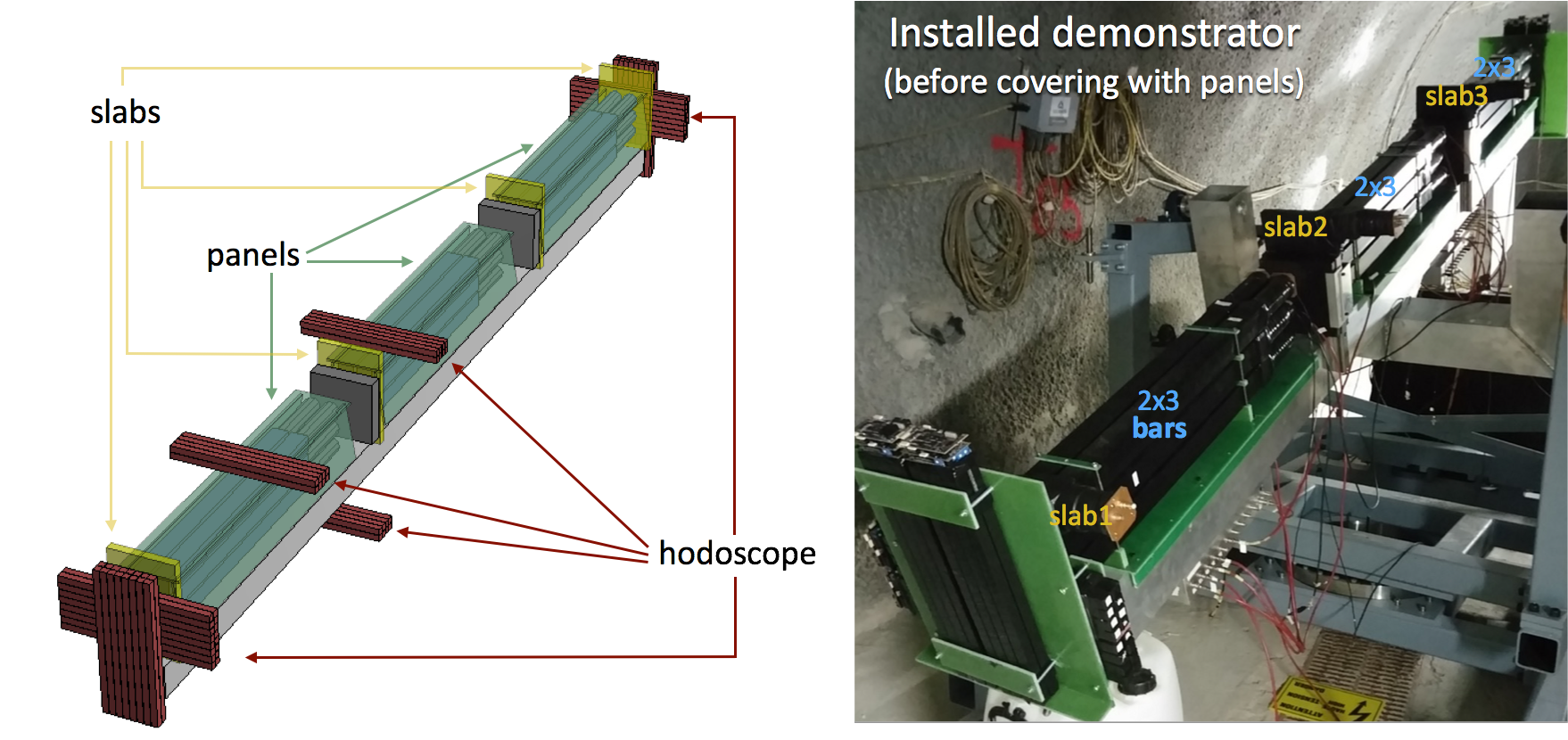}
\caption{Design of the demonstrator and a photo of the demonstrator before covering it with panels.}
\label{fig:demo}
\end{figure}
In order to verify the feasibility and optimize the design of the experiment thoroughly, a 1 \% of the detector was installed as a "demonstrator" in the fall of 2017. As shown in Figure~\ref{fig:demo}, it is composed of 3 layers of $2\times3$ scintillator+PMT units. In addition to the scintillator bars, additional components were installed to reduce or understand certain types of backgrounds; lead bricks were installed between the layers to shield radiation between layers and scintillator slabs were installed to tag through-going particles, get
time info, and shield radiation, \textit{e.g.}, neutrons. Thin and large scintillator panels were installed to cover the top and the sides and to provide an ability to reject cosmic muons. Lastly, hodoscopes were installed to get the tracks of beam and cosmic muons.  

\section{Results from demonstrator}
The demonstrator has been running successfully since its installation, and provided a valuable opportunity to understand the background and optimize the final design of the full detector.  

It is important to calibrate charge because it tells us how small charge the milliQan detector is sensitive to. The calibration is done by calculating the number of photo-electrons($\textrm{N}_\textrm{PE}$) for cosmic muons where $\textrm{N}_\textrm{PE}$ is extracted by dividing pulse area of cosmic muons by the pulse area of single photo-electrons (SPE). Then, the $\textrm{N}_\textrm{PE}$ is extrapolated to fractional charges by $Q^2$. The SPEs are taken from late pulses and
measured in situ. SPE pulse area measurement was also done on the bench as a validation. The typical value of the measured  $\textrm{N}_\textrm{PE}$ for $Q=1e$ is about $5k$. Taking the difference in flight distance of cosmic muons and through-going muons in the scintillator bars (80 cm/5 cm), the $\textrm{N}_\textrm{PE}$ for through-going muon is approximately $5k\times80/5 = 80k$. This gives $\textrm{N}_\textrm{PE}=1$ for $Q \sim 3\times10^{-3}e$, which is consistent with the full Geant4
simulation results~\cite{Ball:2016zrp}.

\begin{figure}[htp]
\centering
\vspace{1cm}
\subfigure[]{
  \centering
  \label{subfig:lumi}
  \includegraphics[width=0.75\textwidth]{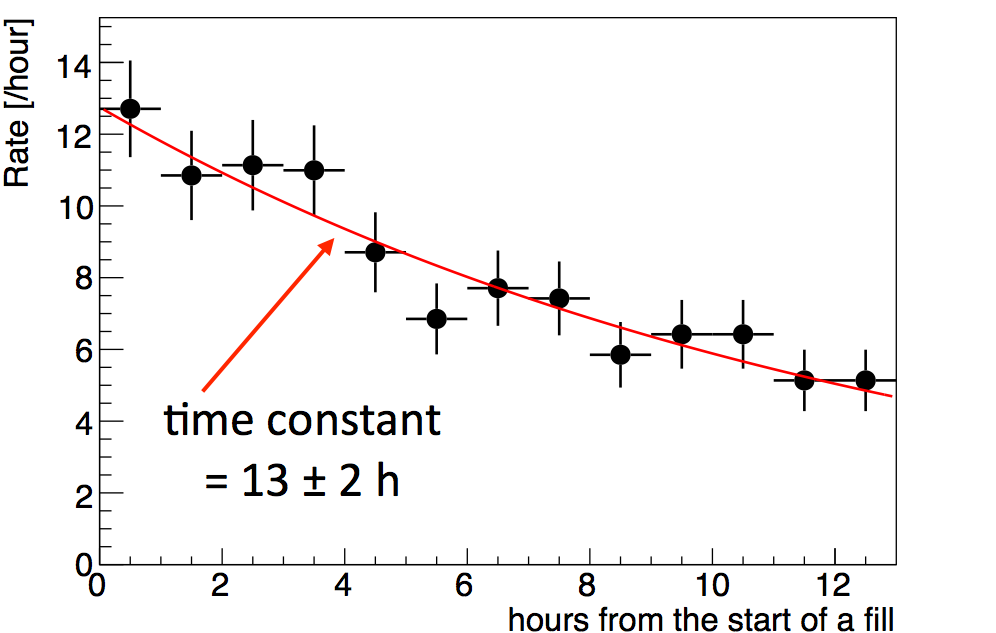}
}
\hspace{1cm}
\subfigure[]{
  \centering
  \label{subfig:time}
  \includegraphics[width=0.75\textwidth]{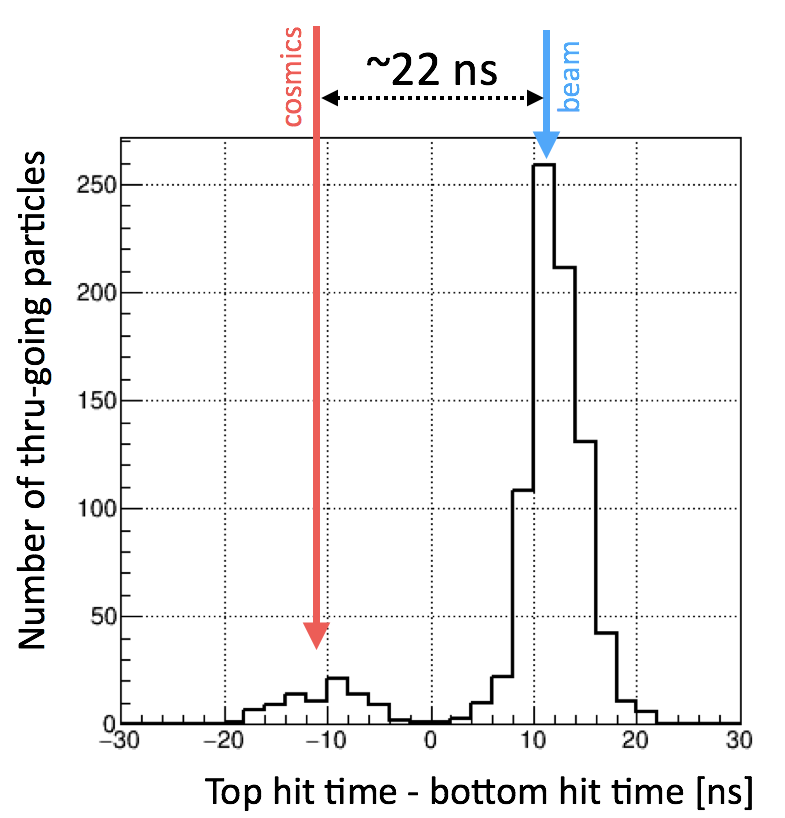}
}
\caption{ (a) Number of through-going particles as a function of time [hour] and (b) the time difference between the top and the bottom slabs. }
\label{fig:lumi_time}
\end{figure}

Using muons from collisions we can understand the alignment, triggering, and timing calibration of the demonstrator. Figure~\ref{fig:lumi_time} shows the rate of beam muons as a function of time for a few LHC fills. The rate decreases exponentially, being consistent with the trend of the instantaneous luminosity. The time constant from the exponential fit is $13\pm2$ hours while that of the instantaneous luminosity is $14$ hours. This is an evidence that the detector "sees" the particles from collisions which indicates the trigger works and the detector is pointing to CMS IP.  

The distance between the top and the bottom slabs is 3.6 m. Therefore, the time difference between the beam muons that hit the bottom slab first then the top one and the cosmic muons that his the top slab then the bottom one is $3.6 \textrm{m} / 0.3 [\textrm{m}/\textrm{ns}] \times 2 = 24$ ns. Figure~\ref{fig:lumi_time} shows the time difference between two slabs. There are two populations peaking at -10 ns and 12 ns. The time difference is consistent with the muons coming from the IP and the sky; the peak around at -10 ns is due to cosmic muons and the peak around at 12 ns is due to beam muons. This shows that the time is measured correctly. 

\section{Summary}
The milliQan experiment can probe the region of $Q=10^{-1}e - 10^{-3}e$ for masses in the range $0.1 - 100$ GeV, unexplored by previous experiments. The detector is composed of 3 layers of long plastic scintillators that produce enough photo-electrons for incoming charged particles with $Q=O(10^{-3}e)$ to be observed. It will be installed in a tunnel at LHC P5, 33 m away from the CMS interaction point. In order to understand the backgrounds more thoroughly, a 1\% version of the detector "demonstrator" was installed and has been running since fall 2017. The data shows that the muons from collisions go through the demonstrator, the detector is sensitive to charges down to $3\times10^{-3}e$, and the cosmic and beam muons can be clearly distinguished using time information. The full scale detector is expected to be installed during 2019-2020 and will provide an unprecedented sensitivity to the uncovered region.    

\section*{Acknowledgement}
This work was supported by the US Department of Energy under award number DE‐SC0011702.

\end{document}